\documentclass[acmsmall,screen,nonacm]{acmart}
\AtBeginDocument{%
  }

\acmBooktitle{ }
\begin{document}

\title{Towards Decentralized Trust and Governance for SSI-Compliant EUDI Wallets using Trusted Execution Environments and  Zero-Knowledge Proofs}

 
\author{Nacereddine Sitouah}
\affiliation{%
 % \institution{Independent Researcher}
  \institution{Polytechnic University of Milan}
  %\city{Milan}
  \country{Italy}
}
\email{nacereddine.sitouah@polimi.it}

\author{Francesco Bruschi}
\affiliation{%
  \institution{Polytechnic University of Milan}
  %\city{Milan}
  \country{Italy}
}
\email{francesco.bruschi@polimi.it}

%\affiliation{%
% \institution{Rajiv Gandhi University}
 %\state{Arunachal Pradesh}
 %\country{India}}

\renewcommand{\shortauthors}{Sitouah et al.}

\begin{abstract}

 The European Digital Identity Wallet (EUDI Wallet), and its Italian implementation, the IT-Wallet, introduced under eIDAS 2.0, enables citizens to store and manage their digital identity and personal attributes through a mobile application. Italy is among the leading European countries in the adoption and implementation of eIDAS, initially through the Carta d'Identità Elettronica (CIE) and Sistema Pubblico di Identità Digitale (SPID) identity schemes, and currently through the development of the IT-Wallet.

Decentralized Identity, also referred to as Self-Sovereign Identity (SSI), is an Identity Management System (IDMS) that leverages Distributed Ledger Technologies (DLTs) to provide individuals with greater control and sovereignty over their digital identities. The EUDI Wallet is required to ensure privacy, liability, and interoperability in accordance with eIDAS 2.0. To achieve these objectives, the IT-Wallet adopts a centralized trust federation architecture in which credential verification relies on the validation of a trust chain rooted in centralized trust anchors. While this approach prioritizes security, legal accountability, and personal data protection, it does so at the expense of self-sovereignty.

This paper proposes a confidential verifier for IT-Wallet credentials that operates within a Trusted Execution Environment (TEE) and is assisted by smart contracts. The proposed architecture enables the integration of IT-Wallet credentials and services into an SSI-compliant ecosystem while introducing a decentralized governance layer based on Shamir's Secret Sharing (SSS) and enforced through publicly auditable smart contracts.
 
\end{abstract}



\begin{comment} 
\ccsdesc[300]{Security and Privacy~Security services, Distributed systems security} 

\ccsdesc[100]{Information systems~Information storage systems}
\end{comment}


%\received{15 June 2026}

%%6
\maketitle

\section{Introduction}

The Electronic Identification, Authentication and Trust Services (eIDAS) Regulation \cite{eidas2014} establishes a common legal framework for electronic identification and trust services across the European Union. eIDAS 1.0 aimed to foster a European digital market by enabling cross-border interoperability of electronic identities (eIDs) and trust services.

However, several shortcomings emerged during its adoption. Differences in national regulations and identity schemes limited interoperability, as Member States often relied on their own certification and assurance frameworks despite the existence of EU-level standards. Furthermore, the complexity and rigidity of the regulation hindered the participation of private-sector service providers and restricted the use of eIDs beyond their original scope. As a result, the integration of digital identity services across Europe remained fragmented. In addition, eIDAS 1.0 primarily viewed digital identities as electronic equivalents of traditional identity documents, overlooking their broader potential. Modern digital identities can incorporate attributes, claims, verifiable credentials, and signed attestations, enabling a wider range of digital services and interactions than conventional identification mechanisms. The above-mentioned limitations have led to a major review and amendment of the eIDAS Regulation, where most drawbacks  have been addressed in order to achieve interoperability in the EU market and include the private and business sectors.

 The revised regulation, eIDAS 2.0 \cite{eidas2024}, addresses many of the limitations of eIDAS 1.0 through the introduction of the European Digital Identity Wallet (EUDI Wallet). The EUDI Wallet enables citizens to securely manage and share digital identity credentials, attributes, and attestations across the European Union. By establishing a common framework for digital identity, eIDAS 2.0 aims to enhance cross-border interoperability, strengthen trust among Member States, and facilitate the integration of both public and private digital services \cite{schwalm2022eidas}. The framework supports both public and private wallet implementations, allowing Member States and trusted providers to offer interoperable solutions that comply with common EU technical and security standards.

 Digital Identity Management Systems (IDMS) provide comprehensive infrastructures and services that enable users to maintain a seamless online presence by supporting core functions such as identification, authentication, and authorization for access to digital resources \cite{laurent2015digital}. A robust IDMS is therefore essential for ensuring trust, security, and usability in interactions between individuals, organizations, and service providers. A digital identity (or electronic identity, eID) is defined as the electronic representation of a person or entity \cite{sedlmeir2021digital}, establishing a relationship between an individual or organization and their digital presence, which may include multiple credentials, accounts, attributes, and entitlements. Digital identities are used to respond to authentication and identification requests from service providers, enabling controlled access to digital services. It is important to distinguish between a user and a digital identity: a user is the real-world entity (human or machine) that operates digital resources, while a digital identity is the set of credentials assigned to that entity. Actions performed under authentication are attributed to the corresponding user.

Self-Sovereign Identity (SSI), also referred to as decentralized identity, represents the latest evolution of digital identity management. Unlike traditional and federated identity models, which rely on centralized identity providers and trusted intermediaries, SSI leverages Distributed Ledger Technologies (DLTs) to enable individuals to manage and control their own digital identities \cite{zwitter2020digital,1,9858139}. The primary objective of SSI is to provide users with greater autonomy over their personal data, allowing them to determine how, when, and with whom identity information is shared. In this model, users are no longer dependent on a single identity provider and can maintain multiple digital identities tailored to different contexts and use cases.

SSI ecosystems are typically built around Decentralized Identifiers (DIDs) and Verifiable Credentials (VCs). DIDs are unique identifiers controlled by their owners and resolvable to metadata such as public keys, while VCs are cryptographically signed attestations issued by trusted entities, such as governments, universities, or healthcare providers. Together, these technologies enable secure, privacy-preserving, and verifiable digital interactions. Credentials are stored in digital identity wallets under the holder's control and can be selectively disclosed to minimize the exposure of personal information while preserving authenticity and integrity.

\begin{figure}[ht]
\centering
%[width=\linewidth]
\includegraphics[width=11.3cm]{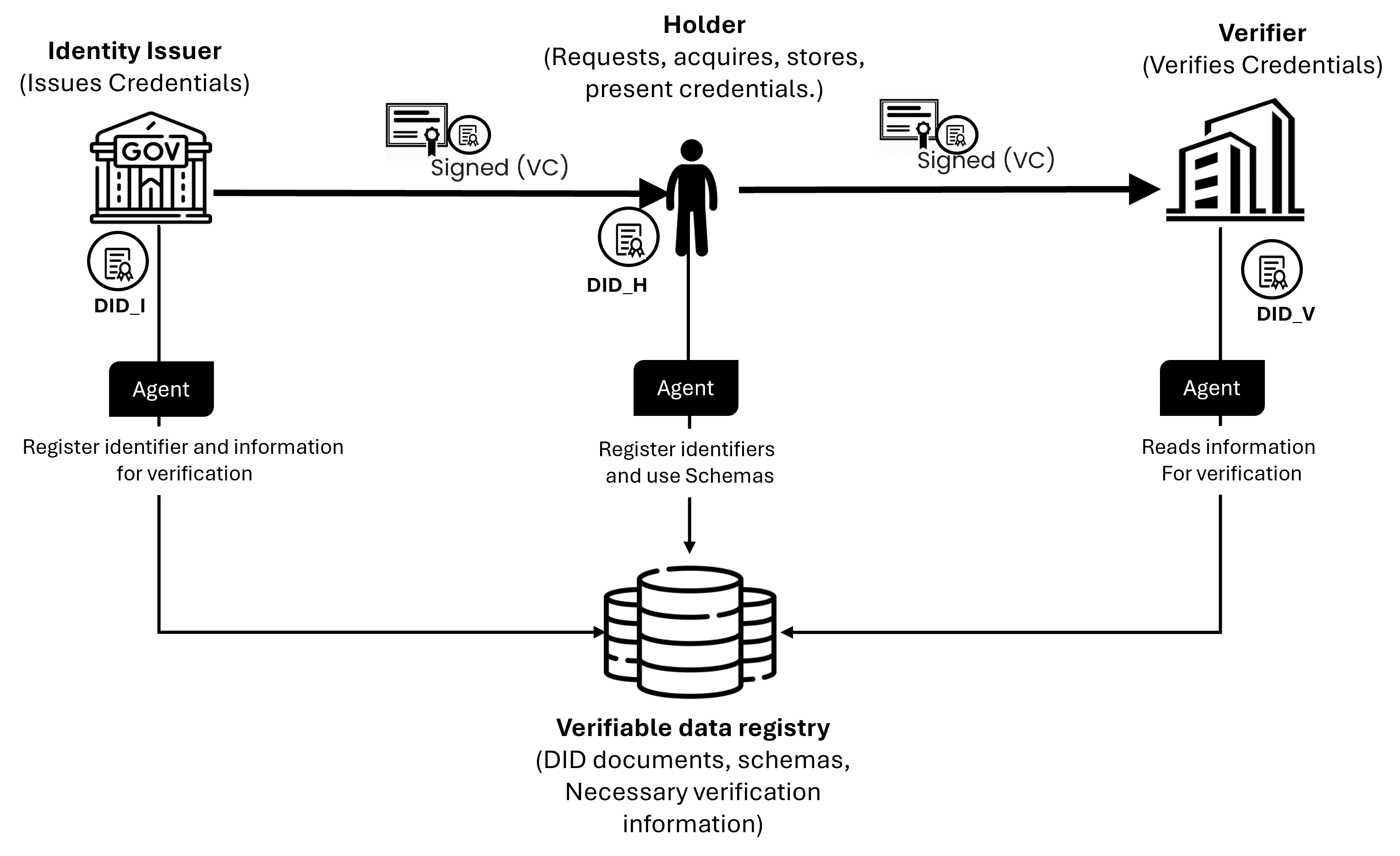}
\caption{ Overview of a typical SSI interaction}
\Description{The main actors in an SSI system publish their DIDs and interact via a Decentralized transparent Storage (i.e. a DLT, Blockchain, etc) in order to verify the validity of a credential. The Identity issuer issues a signed VC to a User's Wallet, which is safeguarded using a recovery "seed phrase". The user then presents the signed VC to a relaying party, which in turn verifies its validity by verifying the signature of the Identity Issuer DID, available on the decentralized transparent storage}
\label{Overview of a typically SSI interaction}
\end{figure}

Credential exchange in SSI relies on a trust model involving three primary roles: the issuer, the holder, and the verifier. Issuers create and sign credentials, holders store and present them, and verifiers validate their authenticity before granting access to a service or resource. Verification is performed cryptographically using publicly available issuer information, often maintained through a Verifiable Data Registry (VDR). This architecture eliminates the need for direct communication with the credential issuer during verification, improving privacy, interoperability, and user sovereignty while maintaining trust in digital transactions---as shown in Fig. \ref{Overview of a typically SSI interaction}.

Multiple structural and regulatory challenges currently hinder private implementations of EUDI Wallets from fully achieving SSI compliance.

\begin{enumerate}
\item \textbf{Centralization of OpenID Connect and limited DLT integration:} Although decentralization is not explicitly prohibited, the Architecture Reference Framework (ARF) and current Member State implementations largely favor OpenID Connect-based architectures built on Public Key Infrastructure (PKI). As a result, status verification mechanisms rely on centralized trust anchors and service endpoints operated by trusted providers to validate credentials, wallet instances, or trust marks. While an EU Blockchain consortium model could improve transparency and accountability, similar objectives may also be achieved through confidential wallet designs that ensure verifiable execution without requiring full reliance on external oversight. However, such approaches require careful integration of privacy-enhancing technologies (PETs), some of which are more naturally supported in consortium-based deployments.

\item \textbf{Restriction to government-issued identifiers:} Within the EUDI Wallet framework, legally recognized authentication is primarily based on government-issued Personal Identification Data (PID), which serve as the authoritative basis for identity verification. Consequently, alternative identifiers are not formally recognized for authentication purposes. This design choice limits the role of privacy-preserving or user-generated identifiers and implies that wallet providers may retain the ability to identify users, thereby constraining full anonymity as envisioned in SSI models.

\item \textbf{Verifier registration requirements:} The regulation requires that relying parties and verifiers of PIDs, Qualified Electronic Attestations of Attributes (QEAA), and other credentials be formally registered. While this enhances accountability and trust, it also introduces operational constraints that may limit openness, interoperability, and the range of feasible use cases for the wallet ecosystem.

\item \textbf{Authentication requirements under eIDAS 2.0:} A key regulatory constraint is that strong authentication is predominantly tied to PID-based identity proofing. Although the Architecture Reference Framework allows the use of non-qualified Electronic Attestations of Attributes (EAAs), these do not provide the same legal assurance level as qualified credentials. This creates a layered trust model in which flexibility is increased at the application level, but formal recognition remains centered on government-issued identity.

\item \textbf{Impact of EUDI Wallet providers:} EUDI Wallet providers are responsible for ensuring the integrity of wallet instances and bear liability for the protection of user data. However, this central role introduces potential risks, including the ability to trace user activity or restrict access without transparency, reflecting limitations typical of centralized PKI-based systems. In addition, providers must manage large volumes of sensitive data, increasing computational overhead and raising concerns regarding data leakage if adequate safeguards are not enforced.

\end{enumerate}

From an SSI perspective, DIDs can provide a strong foundation for authentication. However, to meet high assurance levels (LoA High), identity proofing requirements typically necessitate in-person verification. As a result, practical deployments often link DIDs to government-issued PID during onboarding, balancing decentralization goals with regulatory identity assurance requirements. These challenges make a private EUDIW by design choice, technologically incompatible with SSI.

This article is an extension of our earlier work \cite{sitouah2026enabling}, we extend this approach that explores how EUDI Wallet credentials can be incorporated into SSI ecosystems while preserving user sovereignty. Specifically, \cite{sitouah2026enabling} investigates mechanisms that enable identity holders to generate verifiable proofs independently and examines how such proofs can be represented and verified in a manner consistent with SSI standards and principles. This article also focuses on the origin and security of an SSI-Wallet or workload running under Trusted Execution Environment (TEE), it incorporates a decentralized codebase management framework based on Shamir's Secret Sharing (SSS), MPC threshold signatures and smart contracts The remainder of this paper is organized as follows. Section \ref{Background} introduces the necessary background on TEEs, IT-Wallets, and confidential containers, and reviews related work. Section \ref{solution} presents the proposed SSI-compliant IT-Wallet credential verifier and the decentralized codebase management framework based on SSS and smart contracts. It also describes the cryptographic key management mechanisms supporting workload portability, high availability, and key rotation. Section \ref{Security} defines the threat model and trust assumptions and provides a security analysis of the proposed approach. Finally, Section \ref{Conclusion} concludes the paper by discussing its limitations and outlining directions for future research.

\section{Background}
\label{Background}
\subsection{Trusted Execution Environments}
Trusted Execution Environments (TEEs) provide hardware-backed isolated execution environments that protect code and data from the rest of the system, including privileged software such as operating systems and hypervisors. Unlike traditional hardware security mechanisms that primarily focus on key storage and platform integrity, modern TEE technologies enable the execution of arbitrary workloads within protected memory regions while preserving the confidentiality and integrity of computations. Recent solutions, including Intel SGX and TDX, and AMD SEV-SNP, target different deployment scenarios ranging from mobile devices to cloud infrastructures and confidential computing platforms.

TEEs have become a fundamental building block for applications that require strong security guarantees over sensitive data and computations. They are widely used in secure cloud computing, digital identity systems, cryptocurrency wallets, privacy-preserving analytics, and decentralized applications. By isolating workloads from the underlying infrastructure, TEEs allow users to process confidential information without fully trusting cloud operators, system administrators, or co-located tenants. The choice of a specific TEE technology depends on the desired threat model and security guarantees, including confidentiality, integrity, attestability, and support for secure state recovery.

A key feature of TEEs is their ability to provide cryptographic evidence about the software they execute through attestation mechanisms. Confidentiality ensures that data processed within the TEE remains inaccessible outside the trusted computing base, while integrity guarantees that neither code nor data can be modified during execution. Attestability enables external parties to verify these properties remotely, and recoverability mechanisms allow workloads to securely preserve state across restarts and migrations through protected storage and sealing techniques. Availability, however, typically remains outside the scope of TEE security guarantees and is generally delegated to infrastructure providers.

Trust establishment in TEEs relies on attestation protocols that allow workloads to prove their identity and integrity. Local attestation enables trusted environments running on the same platform to authenticate one another and establish secure communication channels, whereas remote attestation allows external verifiers to validate the authenticity of a workload executing inside a TEE. During remote attestation, a trusted application is measured and executed within the secure environment, after which a dedicated attestation component generates signed evidence containing the workload's measurements and execution state. This evidence, commonly referred to as a quote, is verified against certificate chains rooted in the hardware vendor's trust infrastructure. Through this process, external parties can establish trust in the executed software, securely provision secrets, and verify that computation results originate from an authentic and untampered TEE workload.
\subsection{Confidential containers}

Confidential Containers (CoCo)\footnote{{CoCo : {https://confidentialcontainers.org/}}} is an open-source project hosted by the Cloud Native Computing Foundation (CNCF) that aims to integrate confidential computing into cloud-native environments. CoCo enables containerized workloads to run within hardware-backed TEEs while preserving compatibility with existing Kubernetes workflows  \cite{s1, valdez2024crossing}. The framework abstracts the underlying confidential-computing technology and provides components for workload isolation, attestation, secure image handling, and secret provisioning. A central component of the architecture is the Trustee service, which verifies remote-attestation evidence and ensures that workloads are executing within approved TEEs with the expected software measurements. Following successful attestation, Trustee can securely provision secrets, cryptographic keys, and configuration data only to authorized workloads. Its primary objective is to allow organizations to deploy confidential workloads with minimal changes to existing applications and infrastructure while maintaining strong trust guarantees throughout the workload lifecycle.

Edgeless Contrast\footnote{{Contrast: https://docs.edgeless.systems/contrast/1.9/basics/confidential-containers}} is another confidential-computing platform designed to simplify the deployment and operation of confidential workloads on Kubernetes. In addition to leveraging hardware-backed TEEs, Contrast introduces a comprehensive trust-management layer (the coordinator) that automatically verifies workloads through remote attestation and securely provisions secrets only to attested applications. The framework emphasizes transparency and verifiability by binding deployed workloads to cryptographic measurements and enforcing policies that ensure only approved software receives access to sensitive data. This approach allows organizations to establish trust in cloud-based workloads without relying on a specific trusted authority.

dstack \cite{zhou2025dstack} is an open-source framework for deploying and managing confidential containers across heterogeneous cloud environments. It focuses on simplifying the development and operation of confidential applications by providing integrated support for attestation, secure secret management, and workload deployment. Unlike solutions that are tightly coupled to Kubernetes, dstack offers a more developer-centric abstraction for confidential-computing infrastructures, enabling applications to be deployed across different cloud providers and TEE backends while maintaining a consistent security model. This portability makes it particularly attractive for organizations seeking to avoid vendor lock-in while adopting confidential computing. Compared with CoCo and Edgeless Contrast, which are designed as generic platforms for confidential container deployment, dstack follows a more application-oriented approach. Consequently, adapting dstack to new use cases may require additional integration effort and application-specific modifications compared with the more transparent deployment model provided by CoCo and Edgeless Contrast.

\subsection{ IO IT Wallet and eIDAS 2.0}
The Italian “IO” app\footnote{{ IT-Wallet System --- app IO ({https://ioapp.it/documenti-su-io})}}---often referred to as the Italian Wallet (IT-Wallet)---is the official digital wallet and public services app developed by the Italian government, as part of Italy’s national digital identity ecosystem  and the country’s implementation of the EUDIW initiative under eIDAS 2.0. In its current version, the Wallet's functions are integrated into "IO" app in order to  allows citizens to request and store digital versions of Driver's license, Health card and the European disability card. It grants these digital documents the same legal value of physical cards. IT-Wallet authenticates its users using the existing eID schemes: SPID (Sistema Pubblico di Identità Digitale) and CIE (Carta d’Identità Elettronica).

\begin{figure}[ht]
\centering
%[width=\linewidth]
\includegraphics[width=11cm]{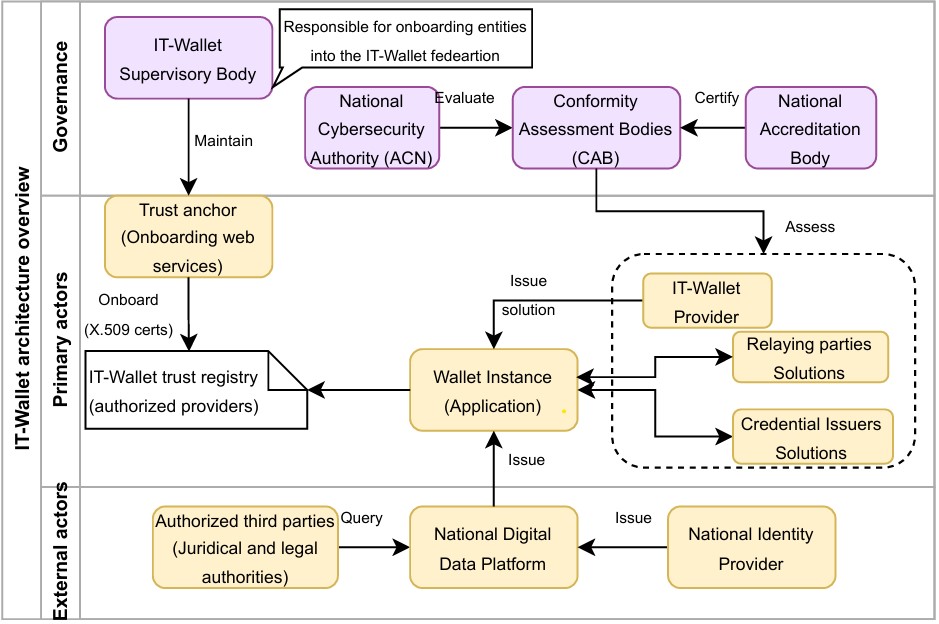}
\caption{ IT-Wallet trust federation architecture: actors interaction \cite{agid2024itwallet,sitouah2026enabling}}

\Description{Diagram illustrating different entities in IT-architecture: Governance bodies, primary actors and external parties}
\label{IT-Wallet trust federation architecture: actors interaction}
\end{figure}

IT-Wallet relies on a Trust Federation (TF) architecture, a multi-layered architecture where governance bodies establish and maintain the trust infrastructure across Italy and the EU. This federation is based on OpenID Connect (OIDC) Federation \cite{openid2024federation1_0} and adopted to meet eIDAS 2.0 Architecture Reference Framework guidelines---as illustrated in Fig. \ref{IT-Wallet trust federation architecture: actors interaction}.  
OIDC is an identity layer built on top of OAuth 2.0 that extends authorization capabilities with user authentication. It enables relying parties to verify a user's identity through an identity provider, which authenticates the user and issues identity-related tokens and claims. In a typical OIDC flow, a user authenticates with an OIDC provider, which subsequently returns an ID Token and Access Token to the relying party, allowing it to obtain verified information about the authenticated user through the UserInfo endpoint.

OIDC Federation (OIDF) extends this model by establishing trust relationships between multiple organizations through a federation framework. Trust is built using signed metadata documents called Entity Statements, which describe federation participants and their properties. These statements are organized into Trust Chains that link a participating entity to a trusted authority, known as the Trust Anchor. OIDF also introduces Trust Marks, which are signed assertions indicating compliance with specific standards or regulatory requirements, and Trust Mark Status endpoints, which allow participants to verify that a Trust Mark remains valid and has not been revoked. Together, these mechanisms enable scalable and interoperable trust management across federated identity ecosystems.

Providers and applications must be vetted in IT-Wallet ecosystem. Certified Conformity Assessment Bodies (CABs) make sure that credential issuers, relaying parties and IT-Wallet provider (private or public) are compliant to eIDAS 2.0 and the corresponding national and EU security standards. An IT-Wallet instance must be added to a trust registry managed by a trust anchor entity that is responsible for on-boarding trusted services. 

 The workflow of a general credential verification in IT-Wallet  requires the verification of Entities Configurations, Entities Statements, Wallet metadata Statement and different Trust Mark JWTs if applicable. These objects and there signatures should be included in the trust chain of a credential, or acquired to rebuild the trust chain and verify its validity. The verification also requires accessing multiple endpoints: Subordinate list endpoint, Trust Mark status endpoint and Wallet attestation status endpoint; these web endpoints return a plain JSON response indicating the status of the subject it attests.

\subsection{Related Work}

%\cite{sharif2024protecting} focus on threat modeling 
%identified that the generic aspect of its framework security and privacy issues in the EUDI wallet framework, and propose a threat modeling desinged to identify potential future threats and controls for secure implementation. 

%outline a systematic threat modeling method, to propose a threat model as a reference tool for practical guidance in the EUDIW ecosystem. 

Previous work explored SSI potential for European electronic identities; \cite{sharif2024protecting} emphasizes that the generic nature of the EUDIW framework may introduce security and privacy risks. The authors apply a systematic threat-modeling methodology to develop a comprehensive threat model that can serve as a practical reference for guiding the development of the EUDIW ecosystem and identifying emerging threats. Furthermore, several studies \cite{konstantinidis2026paradigm, feulner2025self, blasco2025digital, bistarelli2025analyzing} argue that existing digital-identity solutions often diverge from the user-centric principles they seek to promote once deployed in practice. These works highlight persistent challenges related to scalability, interoperability, and the effective preservation of user control over digital identities.

\cite{ramos2025regulatory} explores how ZK proofs can be integrated into the EUDIW ecosystem through approaches that remain compliant with both GDPR requirements and the provisions of eIDAS 2.0 that do not rely on DLTs. The authors discuss the adoption of ZK-based credential issuance, wallet software extensions, and privacy-preserving standalone wallets. Similarly, \cite{zafeiropoulou2025age} demonstrates how the EUDI framework can incorporate privacy-enhancing technologies while satisfying diverse regulatory requirements.

The literature also reflects growing interest in integrating government-issued digital identities with Web3 technologies, although not necessarily from the perspective of decentralized identity management. For example, \cite{omote2025can} proposes a crypto-asset wallet architecture that combines eIDs, one-time wallets, and smart contracts while keeping the eID public key hidden during transaction verification on public blockchains. This approach seeks to prevent the misuse of crypto-assets for illicit activities while preserving user privacy. Meanwhile, \cite{biedermann2024systematisation, baldini2024impact} investigate the incompatibilities between the OIDC-based identity protocols adopted under eIDAS 2.0 and emerging Web3 ecosystems. Their analyses suggest that current research trends favor the development of new trust infrastructures tailored to SSI, potentially increasing fragmentation between EUDIW and decentralized identity ecosystems instead of promoting interoperability through existing standards. To address this issue, \cite{biedermann2025aggregating} proposes a privacy-preserving authentication protocol that extends OIDC with Web3 capabilities. However, the proposed approach remains a largely unidirectional bridge and continues to rely on centralized trust anchors, limiting its alignment with SSI principles.

Decentralized software governance and code management have also attracted increasing attention in recent years. \cite{rangarajan2025codeledger, dhariwal2024decentralized} propose CodeLedger, a DLT-based software versioning system that provides tamper-resistant commit histories, transparent auditing, and secure multi-party collaboration. Similarly, \cite{ramakrishnan2024decentralized} presents a decentralized framework for managing GitHub pull requests using Ethereum smart contracts and browser extensions. The proposed protocol relies on community voting, stake-based governance, and incentive mechanisms to reward constructive participation in software review and approval processes.

A particularly relevant solution is presented in SDIKA\footnote{SDIKA: {https://www.sdika.de/en/}}, which introduces a cloud-based SSI agent designed to preserve user privacy and control by keeping identity data stored locally. Nevertheless, delegating identity-related operations to a cloud provider may conflict with both SSI principles and cloud-security assumptions, as it introduces additional trust dependencies. Other works explore alternative approaches, including the use of ZK proofs, privacy-preserving standalone wallets, and hybrid blockchain infrastructures such as EBSI \cite{RamosFernández02092025,fernandez2024evaluation} to support compliance with eIDAS 2.0. Finally, systems such as PADVA and Janus \cite{szalachowski2019padva, kalka2024comprehensive} leverage the TLSNotary\footnote{TLSNotary : {https://tlsnotary.org/} }  protocol to prove the provenance of data obtained from TLS sessions. These solutions enable the notarization of publicly accessible data by executing TLS communications through privacy-preserving MPC protocols, thereby producing verifiable evidence of data authenticity without requiring trust in a centralized notary.

\section{SSI-Compliant Credential Verification Mechanism}
\label{solution}
%\subsection{Problem description}

%\subsection{SSI-Compliant Framework for IT-Wallet Credential Verification}
We propose a solution that extends our previous work in \cite{sitouah2026enabling}\footnote{\textcolor{blue}{(to replace with IEEE citation once available)}}. A mechanism that leverages IT-Wallet credentials to enable their use within an SSI ecosystem. This article completes the proposed verifier \cite{sitouah2026enabling} with a transparent code governance framework, inspired by dstack Zero Trust Framework \cite{zhou2025dstack} which involves decentralized code management leveraging  smart contracts. 

One of the objectives of the proposed solution, is to make use of IT-Wallet credentials in accordance with its existing design and implementations. Without requiring changes and additional functionalities that are not planned in the design of IT-Wallet on "IO-app". To achieve these objectives, the following is assumed:\vspace{0.15cm} \\ 
a) The proposed architecture relies on the ability to transfer IT Wallet credentials, represented as SD-JWT-VCs, from the IT Wallet to an SSI wallet. At present, the IO app does not support such functionality, as credential export is restricted for security reasons, including the prevention of identity theft, impersonation, and unauthorized credential use. As a result, credentials remain bound to the issuing wallet, limiting portability across wallet ecosystems. However, given the emphasis placed by eIDAS 2.0 on interoperability and cross-border compatibility, support for credential portability may become a desirable feature in future EUDI Wallet implementations. \vspace{0.15cm}
\\b) 
The proposed architecture assumes the use of credentials whose validity is not dependent on second-level temporal accuracy. In other words, a credential verified as valid at time ($t_1$) is expected to remain valid for a sufficiently long interval, such as days or weeks, without requiring continuous status checks. As a result, the approach is not suitable for short-lived credentials, such as Wallet Unit Attestations (WUAs), which are intentionally issued without revocation mechanisms and must instead be validated based on their current validity state at the time of verification. \vspace{0.15cm}
\\c) To establish trust in the proposed credential verification mechanism, the architecture relies on a publicly verifiable permissioned blockchain operated by a consortium of important entities (developers, security auditors, government conformity assessment bodies,  community representatives, etc). As the code governance  framework of the credentials verifier is responsible for validating, approving, and governing the code executed within TEEs, ensuring that only audited and authorized workloads can process IT-Wallet credentials. Through smart contracts and transparent governance procedures, the assumed participating entities can collectively manage software updates, security policies, and attestation requirements while maintaining an immutable and publicly auditable record of all governance decisions. This approach reduces reliance on a single authority, strengthens accountability, and enables independent verification that credential validation is performed by approved code operating within a TEE.

%c) The code governance framework relies on the availability of an immutable consortium }

\subsection{IT-Wallet Credential Verifier Architecture and Workflow}%: Code level isolation}

As presented in \cite{sitouah2026enabling}, a simpler approach to running an IT-Wallet verifier is to execute the verification workload within a code-isolated TEE. In this model, TEE-attested proofs are generated from IT-Wallet---exported credentials by an isolated SSI-Wallet verifier running in a protected memory region, such as those provided by Intel SGX or Keystone (RISC-V). As illustrated in Figure. \ref{Creation of TEE-attested credentials and publication of their succinct ZK  proof}, the generation of an SSI-compliant proof proceeds as follows:

\begin{figure}[h]
\centering
%[width=\linewidth]
\includegraphics[width=12cm]{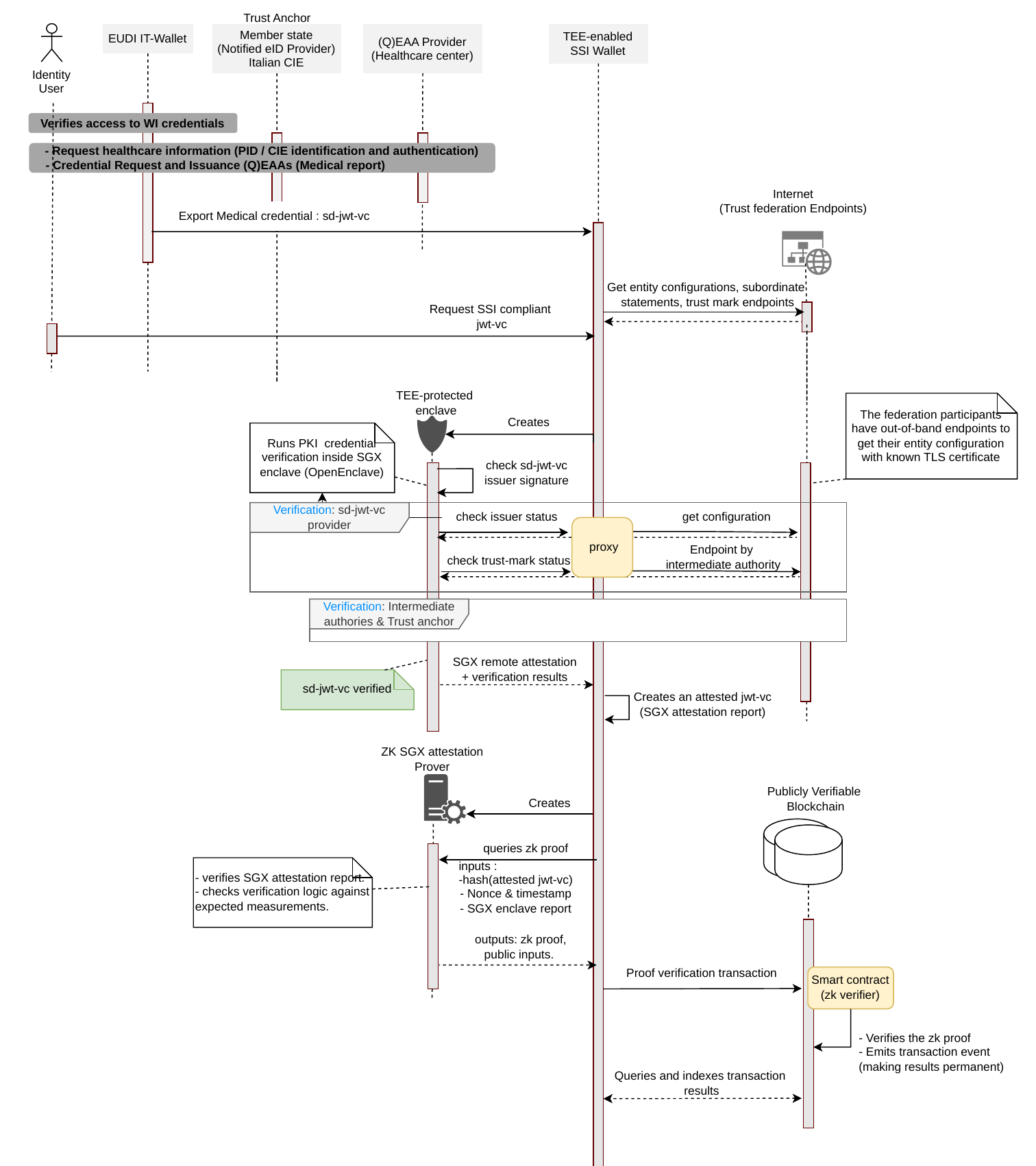}

\caption{Creation of TEE-attested credentials and publication of their succinct ZK  proof}
\label{Creation of TEE-attested credentials and publication of their succinct ZK  proof}
\Description{A sequence diagram describing the simple method of IT-Wallet credentials verifier with code-level TEE isolation and ZK proofs}

\end{figure}
 
\begin{enumerate}

\item Using their Italian eID credentials (SPID or CIE), a citizen accesses an IT-Wallet instance and requests a healthcare credential (QEAA). After successful authentication and authorization by the healthcare provider, the latter issues a medical credential in the form of an SD-JWT-VC. As assumed in the previous section, the IT-Wallet exports the credential as a portable object or file.

\item The TEE-enabled SSI Wallet initiates a preliminary verification of the credential's validity to avoid allocating enclave resources to invalid credentials. This process involves retrieving and validating the relevant entity configurations, subordinate statements, and trust mark endpoints. Once the trust chain has been established and all signatures have been successfully verified, the wallet checks both the trust mark status of the IT-Wallet instance and the credential's revocation status.

\item The SSI Wallet then initializes a secure enclave to perform a trusted verification of the original SD-JWT-VC, together with its associated timestamps and predefined policy constraints. This verification is repeated within the enclave to provide cryptographic assurance that the evaluation is executed in an isolated and trusted environment. Communications between the enclave and external web endpoints are protected using an enclave-enabled TLS library (e.g., mbedTLS-SGX\footnote{{https://github.com/bl4ck5un/mbedtls-SGX}} or OpenSSL-SGX\footnote{{https://github.com/sparkly9399/SGX-OpenSSL}}). Such libraries allow the TLS handshake to be initiated and executed from within the enclave, while the host operating system merely relays network traffic through its network interface. Consequently, sensitive cryptographic material, including TLS session keys and verification state, remains confined to the enclave. This design mitigates threats arising from a compromised operating system or hypervisor and reduces the risk of server impersonation attacks resulting from key theft or malicious manipulation of the verification process.

\item Upon successful completion of the verification process, the enclave produces a hardware-attested proof containing the original credential, verification timestamps, endpoint certificates, and other relevant metadata. Endpoint certificates are included as verification inputs to reduce reliance on external name-resolution mechanisms and mitigate attacks such as DNS poisoning. The SSI Wallet then generates a new JWT-VC credential attesting that the digest of the original credential was successfully validated within an SGX enclave instance executing the expected verification logic.

\item Verifiers can validate the SGX attestation evidence through a certificate chain rooted in Intel's Provisioning Certification Key (PCK). This process preserves execution confidentiality, including with respect to Intel, since the enclave-generated private key never leaves the hardware-protected execution environment. However, repeatedly verifying SGX attestation evidence can be computationally expensive and may introduce unnecessary latency. To address this limitation, the SSI Wallet leverages a zero-knowledge SGX attestation prover. The prover takes as input Intel's root-of-trust certificate, the hash of the newly issued credential, the enclave measurement, and associated attestation metadata. It verifies the SGX attestation evidence and confirms that the credential was processed by the expected enclave code. Subsequently, the prover generates a succinct zero-knowledge proof together with a corresponding smart-contract verifier.

\item The SSI Wallet submits the generated proof to a credential-verifier smart contract deployed on the blockchain. The smart contract validates the zero-knowledge proof and emits an event recording the verification outcome on-chain. Rather than storing the proof and associated metadata in contract storage, the system relies on event logs, which provide the same verifiability guarantees at significantly lower cost while ensuring the verification result remains permanently auditable.  The SSI Wallet subsequently records the verification outcome by either maintaining a local index of the published transaction data---including the credential hash, transaction hash, and block number---or relying on an external indexing service that monitors events emitted by the verification smart contract. Most public Remote Procedure Call (RPC) providers and blockchain indexing services continuously track smart-contract events and retain historical records for extended periods. While delegating indexing to such services simplifies implementation and reduces local storage requirements, it introduces an availability dependency on third-party providers and requires trusting them to maintain complete and accurate historical data.

\end{enumerate}

Although this approach enables credential verification in an SSI-compliant manner, it presents several practical challenges. First, the responsibility for generating TEE-attested credentials falls entirely on end users. In practice, however, not all users possess devices equipped with code-isolated execution  capable of securely running the verification workload. The same limitation applies to the generation of ZK proofs, which may require computational resources and software capabilities unavailable on many consumer devices.

Second, the SSI principle of minimizing reliance on centralized or trusted intermediaries is only partially satisfied. The security of the proposed workflow depends on the correctness and trustworthiness of several software components, including the SSI Wallet, the credential verifier, and the zero-knowledge prover. Consequently, these components become part of the trusted computing base and may introduce additional trust assumptions.

\begin{figure}[h]
\centering
%[width=\linewidth]
\includegraphics[width=\linewidth]{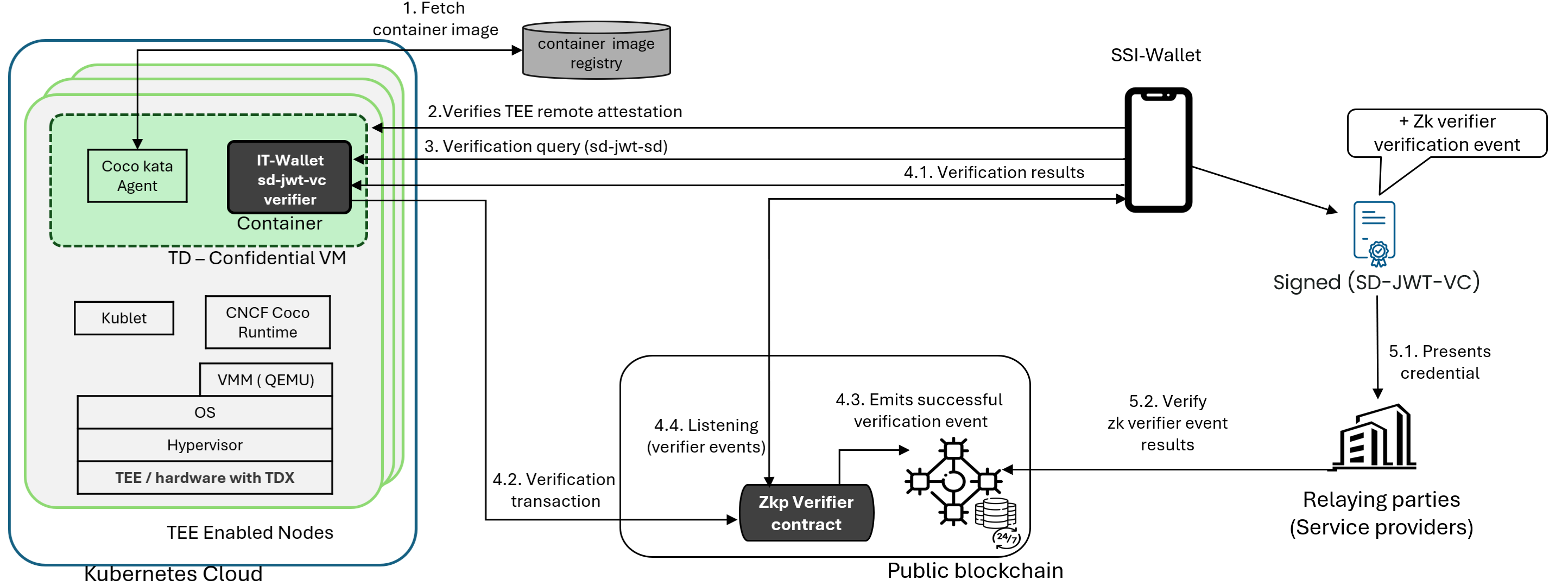}

\caption{Cloud-Based Confidential SSI Wallet and Workflow for Succinct On-Chain Validity Proof Generation}
\label{Cloud-Based Confidential SSI Wallet and Workflow for Succinct On-Chain Validity Proof Generation}
\Description{Workflow diagram of an SSI-Wallet running in confidential containers: 1. The confidential container fetches the SSI-Wallet container image; 2. SSI-Wallet interface verifies the remote attestation of the SSI-Wallet equipped with a credential verifier; 3. SSI-Wallet interface requests verification of an IT-wallet credential; 4. The confidential container returns verification results, publishes a smart contract event on  the ZK verifier contract; 5. SSI-Wallet presents the proof to a relaying party.}

\end{figure}

To alleviate the burden on users and facilitate the integration of IT-Wallet credentials into the SSI ecosystem, the verification and attestation workflow can be offered as a cloud-based service. However, because cloud providers cannot be assumed to be fully trusted with users' identity data, additional privacy-preserving mechanisms are required. To this end, a model based on confidential computing technologies, such as confidential Kubernetes deployments and confidential containers further minimizes the trusted computing base. Frameworks such as CoCo, Edgeless Contrast \cite{s1, valdez2024crossing}, and dstack \cite{zhou2025dstack} leverage hardware-backed TEEs to protect workloads and sensitive data from cloud operators and other privileged infrastructure components. Figure~\ref{Cloud-Based Confidential SSI Wallet and Workflow for Succinct On-Chain Validity Proof Generation} illustrates how confidential containers can be employed to securely perform the verification and attestation of credentials while preserving user privacy.

\begin{enumerate}

\item A cloud provider provisions a confidential-computing platform by deploying a Kubernetes cluster capable of hosting confidential virtual machines (CVMs) or confidential containers. The IT-Wallet credential verification service is packaged as a containerized workload and executed within a TEE-backed execution environment. Upon startup, the CVM retrieves the SSI-Wallet container image equipped with a credential verifier  and generates attestation evidence that cryptographically binds the running software stack---including the container image, runtime configuration, and underlying hardware measurements---to the TEE. To enable independent verification and transparency, the SSI-Wallet source code, build artifacts, and container images should be publicly available and reproducible. 

\item Before importing credential data to the confidential Wallet, the SSI Wallet performs remote attestation of the confidential verifier. Specifically, it verifies the attestation evidence against the hardware vendor's root of trust and confirms that the reported measurements correspond to the expected verifier workload. The attestation also establishes that the workload is executing within a genuine TEE-protected environment and that its memory and execution state are protected from privileged cloud infrastructure components, including the host operating system, hypervisor, and cloud administrators. Reference measurements and software hashes required for validation are made publicly available by the service operator and are cryptographically bound to the attestation evidence.

\item Once the attestation has been successfully verified and the confidentiality guarantees of the execution environment have been established, the SSI Wallet securely transmits the original SD-JWT-VC and any associated verification parameters to the confidential verifier. The communication channel may itself be attested and cryptographically bound to the TEE instance, ensuring that credential data is disclosed only to the verified workload executing within the trusted environment.

\item The verification process performs the complete credential-validation workflow described previously. This includes validating the credential's signature chain (trust chain), resolving and verifying federation metadata, evaluating subordinate statements and trust marks, checking revocation and trust-mark status endpoints, and enforcing any application-specific policy constraints. Upon successful verification, the verifier generates a new attested credential containing a digest of the original credential together with verification metadata and timestamps. The verifier subsequently validates the TEE attestation evidence, generates a succinct zero-knowledge proof attesting to the correctness of the verification process, and submits the proof to the on-chain verifier smart contract. The resulting transaction and emitted event provide a publicly verifiable record that the credential was successfully validated by the expected confidential workload.

\item When the user subsequently presents the original SD-JWT-VC to a relying party, the presentation is accompanied by a reference to the corresponding on-chain verification record, or emitted event identifier. The verifier can independently   confirm that the associated zero-knowledge proof was successfully verified by the smart contract. This approach enables verifiers to rely on a succinct and publicly auditable proof of credential validity without having to repeatedly execute the full federation-validation process themselves or requiring computation resources from IT-Wallet users.

\end{enumerate}

Integrating confidential Kubernetes platforms (e.g., CoCo or Edgeless Contrast) or confidential Docker runtimes (e.g., dstack) addresses the resource-availability limitations of the previously described model while preserving user privacy through hardware-backed isolation. By ensuring that credential-processing workloads execute within remotely attested TEE, sensitive identity data remains protected from cloud operators, infrastructure administrators, and other privileged software components. Consequently, this approach aligns well with the confidentiality and data-minimization principles promoted by eIDAS 2.0. Nevertheless, the deployment and operation of such confidential services introduce several challenges and limitations:

\begin{itemize}

\item \textbf{Confidential container heterogeneity:}
The confidential-computing ecosystem remains highly fragmented. Different platforms rely on distinct TEE technologies, each exposing different attestation formats, trust anchors, security guarantees, and operational constraints. As a result, verifiers and SSI Wallets must support multiple attestation mechanisms and validation procedures to ensure interoperability across providers. This heterogeneity complicates deployment, increases implementation complexity, and may hinder portability of confidential workloads between cloud environments.

\item \textbf{Codebase centralized governance:}
Although confidential computing reduces the need to trust cloud infrastructure operators, users must still trust the software executed within the attested environment. In practice, the verifier workload, federation-validation logic, cryptographic libraries, and zero-knowledge proving components are typically developed and maintained by a limited set of organizations. Consequently, these entities effectively govern the trusted codebase and become part of the system's trusted computing base. Bugs, supply-chain compromises, malicious updates, or governance failures affecting these components could impact all users relying on the service. This introduces a degree of centralization that may conflict with the SSI objective of minimizing reliance on single trusted parties.

\item \textbf{Permanent data loss during downtime:}
Confidential workloads are intentionally designed to minimize trust in the hosting infrastructure and often avoid persistent storage of sensitive user data. Consequently, service interruptions, enclave failures, accidental deletion of sealed state, or loss of cryptographic keys used to protect persistent data may render previously generated verification artifacts unrecoverable. While blockchain records and emitted events remain publicly accessible, auxiliary data such as credential indexes, attestation metadata, cached federation documents, and verification history may be permanently lost unless appropriate backup, replication, and recovery mechanisms are implemented. Designing such mechanisms is challenging because they must preserve confidentiality guarantees while ensuring long-term availability.

\end{itemize}

These limitations suggest that, although confidential computing platforms provide a practical foundation for privacy-preserving credential verification services, additional mechanisms are required to improve interoperability, reduce governance centralization, and guarantee long-term availability of verification artifacts and service state.

%\subsubsection{Architecture Overview}
 
\subsection{Decentralized Private IT-Wallet codebase governance}

Decentralized codebase management can be achieved through a variety of governance and verification mechanisms that distribute trust across multiple independent stakeholders rather than concentrating it within a single software provider. In the context of confidential SSI-compliant Wallet workloads, such approaches enable multiple parties---including wallet developers, security auditors, standards organizations, academic institutions, public authorities, and community representatives---to collectively assess, review, and approve the trustworthiness and security of the software before it is deployed within a TEE.

Several mechanisms can be employed to realize this model. Most commonly, software releases may require threshold-based approval from multiple maintainers through multi-signature code-signing schemes, ensuring that no single entity can unilaterally introduce changes into the trusted codebase. Independent security audits performed by multiple organizations can further strengthen confidence in the software, while reproducible builds allow any party to verify that publicly available source code corresponds exactly to the binaries executed within confidential-computing environments. Additionally, software measurements and approved workload hashes can be published in transparent registries or blockchain-based ledgers, enabling verifiers to independently validate that attested workloads correspond to community-approved versions.

By distributing responsibility for code review, approval, and release management across multiple independent stakeholders, decentralized codebase governance mitigates the risks associated with single-provider control, software supply-chain attacks, malicious updates, and governance failures. More importantly, it aligns the management of trusted SSI-Wallet workloads with the decentralization principles that underpin SSI, reducing reliance on a single trusted software authority. Consequently, trust in the SSI-Wallet ecosystem can be established through a transparent, auditable, and collectively governed process rather than being delegated to a single vendor. Building upon this premise, this paper proposes a decentralized code-governance framework that combines multi-party workload approval,data portability, secure recovery procedures, and cryptographic secret-rotation mechanisms to maintain both the security and long-term resilience of confidential SSI-Wallet services.

The overview architecture of the decentralized code management described in Figure. \ref{Overview of a typically SSI interaction} is as follows: 

\begin{figure}[h]
\centering
%[width=\linewidth]
\includegraphics[width=\linewidth]{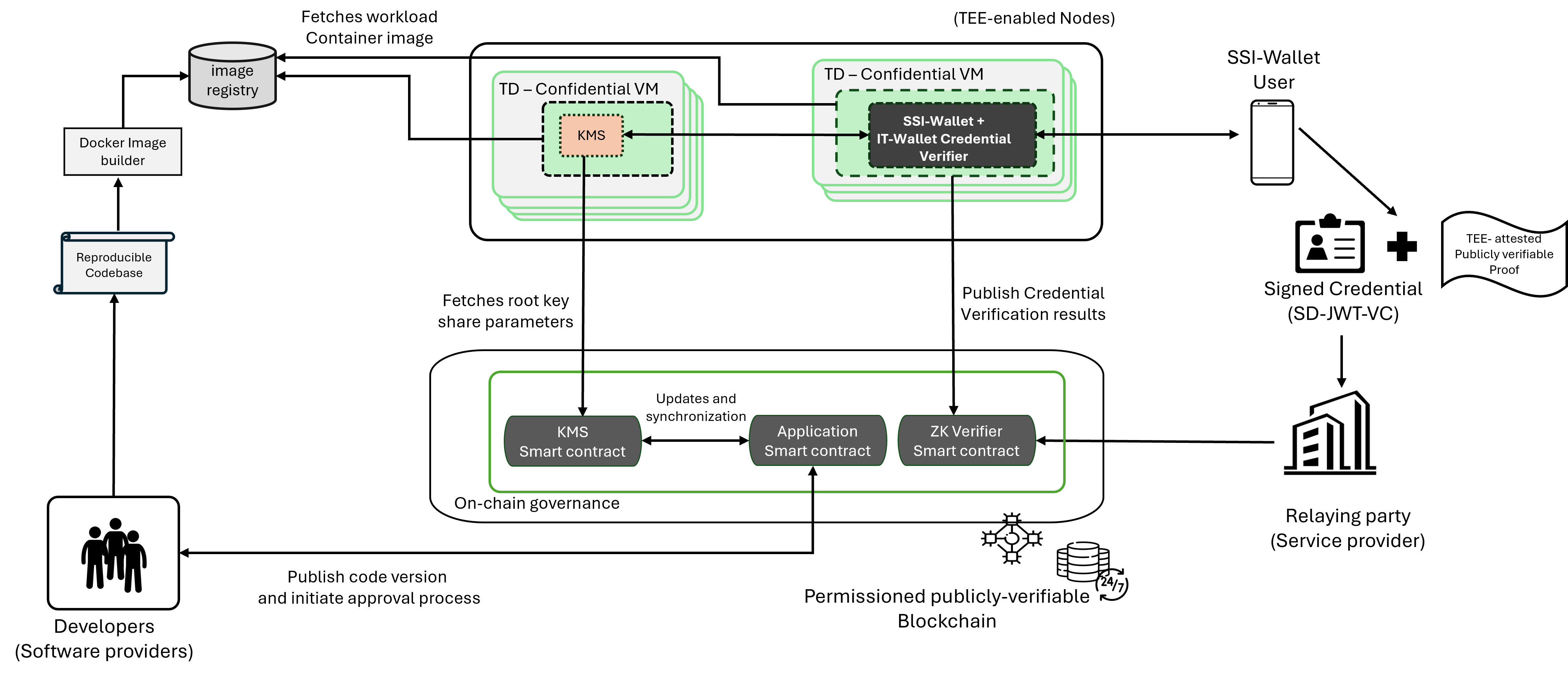}

\caption{Overview of the Decentralized Codebase Management Architecture}
\label{Overview of the Decentralized Codebase Management Architecture}
\Description{ Architecture overview of the decentralized codebase governance, illustrating different actors, TEE instances and smart contracts. }

\end{figure}

\begin{itemize}

\item A community of independent developers maintains the source code and reproducible builds of the SSI Wallet, the IT-Wallet credential verification and zero-knowledge proving service, and a confidential decentralized Key Management Service (KMS) intended to operate within TEEs. To establish a decentralized governance model, an initial set of security auditors and application approvers is defined and registered within the governance framework.

\item An application-governance smart contract manages the lifecycle of each confidential workload. It maintains the registry of approved codebase versions, deployment policies, and code-approval requirements. When authorized developers propose a new workload codebase version, they submit the corresponding codebase hash, build artifacts, and metadata to the governance contract. Security auditors are then notified of the proposal and independently review the source code, reproducible builds, and associated documentation. If the proposed changes satisfy the security, integrity, and confidentiality requirements of the system, auditors approve the proposal by digitally signing its identifier and codebase hash. Once the number of valid approvals reaches a predefined threshold, the proposal is marked as security-approved and becomes eligible for final governance approval.

\item Following the security-audit phase, a separate group of application approvers evaluates the candidate releases that have successfully passed security review. This additional governance layer allows operational requirements to be assessed independently from technical security considerations. Approvers cast their votes by signing the proposal identifier and approved codebase hash. When the required approval threshold is reached, the governance contract designates the corresponding codebase version as the current authorized release and notifies the KMS governance  smart contract of the update.

\item The KMS governance smart contract maintains a registry of authorized applications and deployment policies, including the currently approved codebase versions and expected workload measurements for both the confidential KMS service and the SSI Wallet workloads. Upon receiving a notification from the application-governance smart contract, the KMS smart contract updates its authorization records to reflect the newly approved workload versions. Consequently, only confidential workloads executing approved codebases and producing the corresponding attestation measurements become eligible to access protected cryptographic material.

\item The confidential decentralized KMS is operated by a set of decentralized nodes executing within TEEs. These nodes collectively hold shares of a secret generated using a threshold secret-sharing scheme. Together, the shares protect an application Master key that is never stored or exposed in its entirety by any individual node. Access to key shares is restricted to SSI Wallet instances that satisfy the deployment and attestation requirements defined by the KMS governance contract. When an SSI Wallet requests access to protected secret shares, it establishes an attested TLS channel with the KMS nodes and provides remote-attestation evidence proving its identity and software measurements. Each KMS node independently verifies that the received attestation matches the approved workload measurements recorded on-chain before releasing its corresponding key share.

\item Once the SSI-Wallet collects a sufficient number of valid key shares, it reconstructs the application Master key within its confidential execution environment. This master key is subsequently used to derive application authentication keys, encryption keys, and other cryptographic material required by the wallet. Because identical codebases derive keys from the same Master secret, independently deployed SSI-Wallet instances can securely share cryptographic identity and encrypted state, thereby improving availability, simplifying synchronization, facilitating disaster recovery, and enabling seamless migration between confidential-computing environments.

\end{itemize}

\subsection{Key management Service: Nodes and Smart contracts}

The KMS nodes form a TEE-assisted peer-to-peer network. They use Attested-TLS (Edgeless Contrast) or ZeroTrust-TLS (dstack) to establish secure communication channels, verify attestations, exchange secrets, and maintain service availability. Together, these nodes constitute an off-chain peer-to-peer infrastructure. Initially, the first KMS node—or an initial group of KMS nodes—is manually registered in the application smart contract and serves as the bootstrap trust anchor for the network. This bootstrap node is responsible for verifying and admitting subsequent KMS nodes before they can participate in key-management operations and network services.

Trust decisions are collectively enforced through smart contracts, decentralized governance mechanisms, and cryptographic proofs. In such an architecture, the blockchain serves as the authoritative source of truth for workload authorization, membership management, and deployment/attestation policies, while confidential-computing components provide execution confidentiality and integrity.

In order to dispense of the need of a trustee or coordinator, that verifies new KMS nodes, every authorized node has the capability of verifying the attestation and measurements of subsequent KMS nodes. When a new confidential workload or KMS node wishes to join the network, it first deploys within a TEE and generates remote-attestation evidence containing its workload measurements, platform identity, and execution state. Rather than submitting this evidence to a privileged coordinator or trustee, the node publishes a registration request to the application  smart contract. The request contains the workload identifier, attestation evidence, network endpoint information, and any metadata required to establish communication with other participants. The smart contract records the request and emits an event notifying existing network participants of the pending admission request.

Normally, existing authorized participants independently retrieve and verify the attestation evidence. Verification includes validating the certificate chain associated with the TEE manufacturer, checking the authenticity of the attestation signature, confirming that the reported measurements correspond to an approved workload version, and ensuring compliance with the deployment policies defined on-chain. Each participant then records its approval by submitting a signed vote to the application contract. Once the number of valid approvals reaches a predefined threshold, the application smart contract automatically accepts the new participant and updates the network membership registry (KMS smart contract). This process removes the need for a centralized admission authority while ensuring that trust decisions are collectively governed by multiple independent entities.

An alternative way replaces individual voting with succinct cryptographic proofs. In this model, one or more attestation verifiers validate the remote-attestation evidence off-chain and generate a ZK proof attesting that the verification process was executed correctly according to a predefined policy. The proof demonstrates that the workload's measurements match an approved codebase version and that the attestation chain terminates at a trusted hardware root (Vendor public key) without revealing the complete attestation evidence. The application smart contract verifies only the succinct proof. Upon successful proof verification, the workload is automatically admitted into the network.

Once admitted, the newly authorized KMS node becomes eligible to participate in decentralized key-management operations. Existing KMS nodes periodically retrieve the membership registry maintained by the KMS smart contract and establish secure communication channels only with approved participants. Secret distribution, threshold key resharing, and key-recovery procedures are performed exclusively among nodes whose identities and measurements satisfy the on-chain authorization policies (authorized via the application smart contract). Consequently, the blockchain acts as a decentralized trust anchor governing network membership, while confidential-computing technologies provide execution confidentiality and cryptographic attestations of software integrity. Figure. \ref{gistration of TEE-assisted KMS node} presents a sequence diagram illustrating the complete workflow of decentralized membership management approach, including workload registration, attestation verification, decentralized approval, on-chain membership management, and the subsequent establishment of trust relationships between authorized confidential workloads and KMS nodes. The successful registration flow for a new KMS node is as follows:

\begin{figure}[h]
\centering
%[width=\linewidth]
\includegraphics[width=\linewidth]{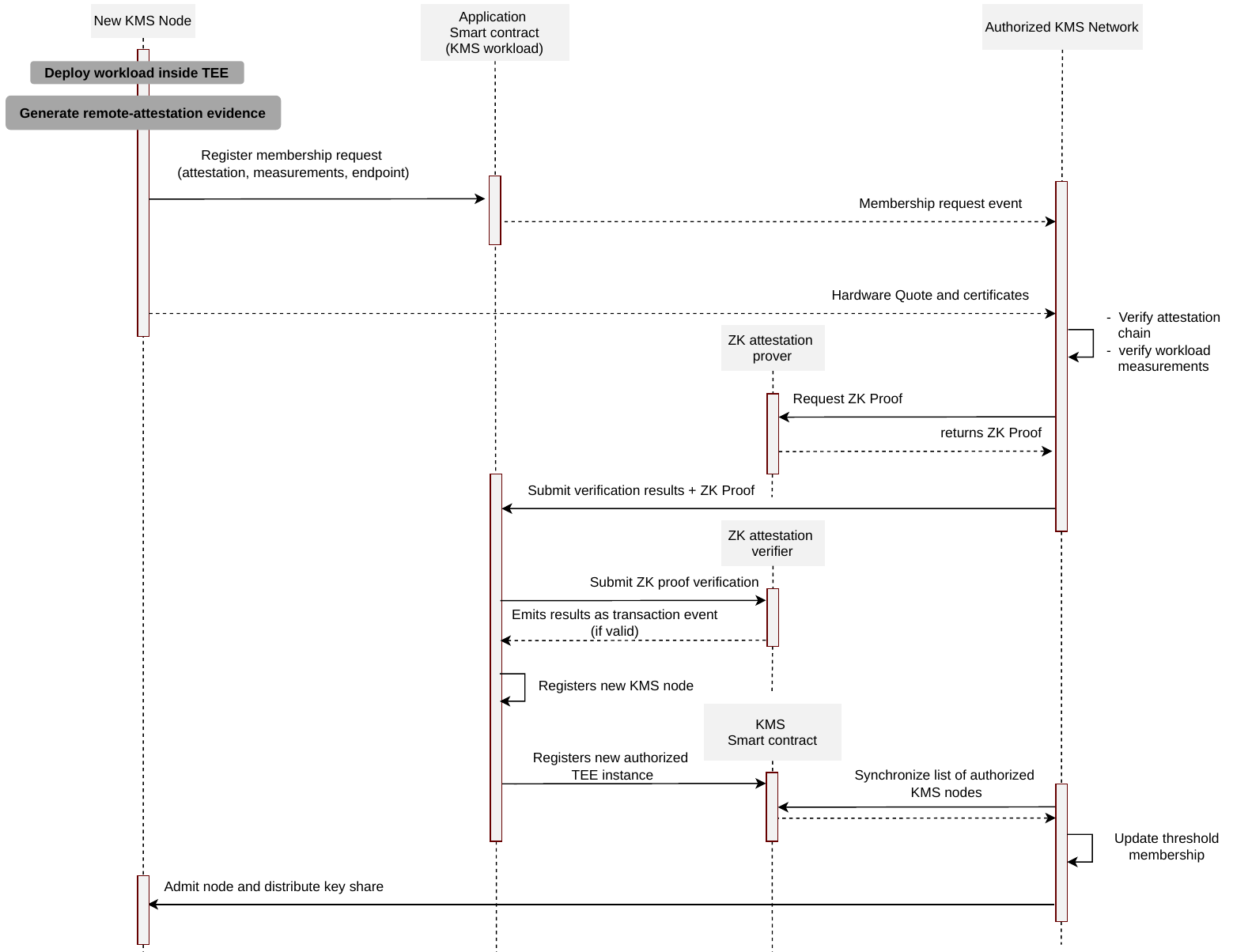}

\caption{Registration of TEE-assisted KMS node}
\label{gistration of TEE-assisted KMS node}
\Description{A sequence diagram describing the registration of a KMS node via TEE attestations verification and smart contracts.}

\end{figure}
\begin{enumerate}
    \item A new KMS node is deployed inside a TEE and generates remote-attestation evidence containing its workload measurements, platform identity, and network endpoint information.

    \item The node submits a registration request to the application smart contract, which records the request and emits a membership-registration event to the authorized KMS network (Nodes already authorized and verified).

    \item Authorized KMS nodes retrieve the attestation evidence and independently verify the hardware quote, certificate chain, and workload measurements against the approved KMS workload specification.

    \item Upon successful verification, the KMS network (at least one Node) requests a ZK proof from a ZK attestation prover. The prover validates the attestation evidence and produces a proof attesting that the workload satisfies the required security and deployment policies.

    \item The KMS network submits the verification results and corresponding ZK proof to the application smart contract.

    \item The smart contract invokes a ZK attestation verifier smart contract to validate the submitted proof. If the proof is valid, the verification result is permanently recorded through a transaction event.

    \item Following successful verification, the application smart contract registers the new KMS node and notifies the KMS smart contract of the newly authorized confidential workload.

    \item The KMS smart contract updates its registry of authorized TEE-assisted KMS instances. Existing KMS nodes synchronize the updated membership list and adjust the threshold secret-sharing configuration accordingly.

    \item Once the new node becomes an authorized member of the KMS network, it participates in key-management operations and receives its corresponding key share through a secure attested communication channel.
    
\end{enumerate}

All membership decisions become transparent, auditable, and publicly verifiable through immutable blockchain records. Likewise, codebase updates, governance rules, and deployment policies can be managed through decentralized approval mechanisms. In particular, threshold-based authorization schemes built upon Shamir's Secret Sharing can be employed to distribute decision-making authority among multiple independent participants, providing  fault tolerance, encryption secrets secrecy, and resistance to malicious actors. The following subsection details how such threshold cryptographic mechanisms can be integrated into the governance and key-management processes to achieve decentralized trust management throughout the system lifecycle.
 
%Subsequent KMS nodes, can be anyone in possession of a TEE device, as long as the application smart contracts authenticates and approves its remote attestation (expected hardware measurements---inc. container image hash) . 

\subsection{Decentralized Key Management: KMS smart contract}

The KMS smart contract serves as the authoritative source that determines which TEE applications are authorized to access the Master key. It also governs the transition process during codebase upgrades, whereby a newly approved KMS workload version becomes the expected reference implementation for all participating KMS nodes. Following an update, KMS nodes gradually migrate to the new codebase and converge toward the approved version. During this transition period, both old and new versions may coexist according to the deployment policy. Once the migration is complete, nodes running outdated codebase versions are no longer authorized to participate in KMS operations or communicate with other members of the KMS network until they upgrade. The process governing codebase updates for both KMS workloads and SSI-compliant wallet workloads follows a similar approval and deployment lifecycle.

When a workload is running as expected and governed in the application smart contract, this workload's identity is added the KMS smart contract. This allows KMS nodes to distribute key shares to the workload in order to reconstruct the Master key via a Shamir's Secret Sharing (SSS) protocol.  A KMS workload deploys a key-evolution architecture combining SSS and  hash-chaining. The system is designed to provide distributed trust for key initialization and forward-secure key evolution for credential processing in EUDI Wallet environments.

At system bootstrap, a secret seed $Seed_0$ is generated  through a Distributed Key Generation (DKG) protocol. The seed is never stored in plaintext outside the trusted boundary. The seed is then used to derive the Master key.
No single node can reconstruct the seed; at least $t$ participants are required for reconstruction or rotation events. The scheme involves different types of key rotations: 
\begin{enumerate}
    \item \textbf{Hash-chain step}: this is an epoch rollover that happens at a very low cost, with backwards secrecy. KMS smart contract identifies each Master key with a unique identifier $id_K$; each $id_K$ indexes  a monotonically increasing epoch identifier on the smart contract $Epoch_{id_k}$. This  epoch parameter is part of the application deployment policies, it is set to increment periodically in order to trigger a low cost Master key change.  Given F(x) a Key Derivation Function (KDF), the Master key is generated as follows: 
    \begin{itemize}
        \item $MasterKey_0=F(Seed_0, salt_{system}, epoch_0)$, with $salt_{system}$ a fixed public value.
        \item For each epoch value  set by the KMS smart contract: \\ $MasterKey_i=F(Seed_0, salt_{system}, Hash(MasterKey_{i-1}||epoch_i)$
    \end{itemize}
    \item \textbf{Fresh Decentralized Key Generation  (DKG)}: upon a periodic period triggered by the KMS smart contract, or an emergency event after compromise requested by nodes, a fresh full DKG protocol is initiated. Similarly to how the first $MasterKey$ is first generated, the protocol works by having each KMS node generate a random secret contribution and use it as the constant term of a Shamir's polynomial of degree $t-1$. Each node distributes shares of its polynomial to all other participants and publishes commitments that allow the shares to be verified. After receiving and validating all shares, each KMS node combines the shares received from all participants into different single final share. Collectively, these final shares correspond to a single secret $Seed_i$, that will computable only by authorized TEE instances.

 \item \textbf{Key-share refresh:} this mechanism enables the KMS nodes to periodically generate and distribute a fresh set of secret shares while preserving the underlying shared secret. As a result, the secret reconstruction procedure remains unchanged and continues to require the same $(t,n)$ threshold of participants. Unlike a full DKG protocol, share refresh operations do not create a new secret and therefore incur significantly lower computational overhead. Instead, participants collaboratively update their local shares through a distributed resharing procedure that preserves the original secret while replacing all existing shares---usually by introducing a new polynomial.  From a security perspective, periodic share refresh provides strong protection against gradual compromise attacks that slowly undermine the Master key security. 
    \\Although this share refreshing and DKG both require a communication complexity $O(n^2)$,  its advantage is that it preserves the existing secret and only refreshes the shares, avoiding the costs of generating a new secret, updating public keys, migrating dependent state, and running additional DKG verification steps. As a result, resharing is often cheaper in practice despite having similar communication complexity.   
    %: shares exposed before a refresh become cryptographically obsolete and cannot be combined with newly generated shares to reconstruct the secret. Consequently, an adversary must compromise at least (t) valid shares within the same refresh interval to recover the protected secret, substantially reducing the risk associated with long-term key exposure in a dynamic KMS network.

\end{enumerate}

The Master key is subsequently used as a root secret from which each application derives cryptographic material according to its specific security requirements. Through a key derivation mechanism, applications generate dedicated encryption and communication keys that are used to protect data at rest and data in transit. This design minimizes exposure of the master key, enables cryptographic separation across applications, and facilitates key lifecycle management while preserving a common decentralized trust anchor, in the KMS smart contract and peer-to-peer network .

\section{Security Model and Analysis}
\label{Security}
%\subsection{Trust assumptions}
%\begin{itemize}
%    \item TEE Hardware providers: 
%\end{itemize}
\subsection{Threat Model}
We consider adversaries that may include developers, administrators, cloud providers, control-plane operators, or any party with privileged access to the underlying infrastructure. Adversaries may attempt to deploy malicious container images, spoof TEE attestations, compromise individual (or a none significant number) Key Management Service (KMS) nodes, intercept network communications through attacks such as DNS hijacking or man-in-the-middle attacks, bypass governance mechanisms to perform unauthorized software updates, or execute workloads outside TEEs. We further consider a malicious infrastructure provider capable of controlling the host system on which the TEE executes, including inspecting memory, modifying software outside the enclave, and replaying previously observed states.

We assume that the TEE provides confidentiality and integrity guarantees for code and data during execution, and that remote attestation accurately reflects the identity and integrity of the deployed workload. We further assume that at least $t$ out of $n$ KMS nodes remain honest during secret reconstruction and key release operations. Trust is therefore distributed across multiple parties and anchored in hardware-based attestation rather than a single trusted authority. While adversaries may observe all blockchain data, public commitments, and protocol messages, they cannot recover protected cryptographic material without compromising the threshold of KMS nodes or breaking the underlying cryptographic primitives.

Our threat model excludes side-channel attacks and other hardware-specific attacks against confidential computing technologies. It also excludes availability attacks, including denial-of-service (DoS) and distributed denial-of-service (DDoS) attacks targeting the network, infrastructure, or system endpoints.

\subsection{Security Analysis}
%From a security perspective, periodic share refresh provides strong protection against gradual compromise attacks: shares exposed before a refresh become cryptographically obsolete and cannot be combined with newly generated shares to reconstruct the secret. Consequently, an adversary must compromise at least (t) valid shares within the same refresh interval to recover the protected secret, substantially reducing the risk associated with long-term key exposure in a dynamic KMS network.

\subsubsection{Confidentiality} 
\leavevmode\\
The combination of SSS protocol and TEEs guarantee that data remain confidential inside TEE, and during export/import  between TEE instances. The secret seed $S_i$ cannot be reconstructed unless at least $t$ KMS nodes collude. Even after reconstruction, the seed exists only transiently within an attested TEE, preventing exposure at the infrastructure level. Derived keys generated via the hash chain remain computationally unlinkable across epochs. 

\subsubsection{Forward Secrecy and Mitigation Cryptanalytic Data Collection} 
\leavevmode\\ 
The fresh DKG protocol generates  completely new shares for the secret $Seed_i$. If it happens that the $MasterKey$ is compromised, subsequent encryption remain safe. Although the hash-chain step in the proposed solution provides backward secrecy alone, as its goal is to prevent ciphertext cryptanalysis attacks. To provide forward secrecy, KMS node can agree on a one hash-chain that is only disclosed to authorized TEE instances, but is has a commitment on-chain that must be verified via hashing algorithms. Allowing automatic synchronized key rotation.  

\subsubsection{Infrastructure and Platform Integrity} 
\leavevmode\\ 
TEE remote attestation overcomes the reliance on the integrity of the host OS and container runtime, in which a compromised host can inspect container memory, extract secrets, modify application binaries, or alter execution behavior. However, confidential containers execute with  memory contents that remain encrypted and inaccessible to the host environment. Consequently, privileged infrastructure operators cannot directly access application state, cryptographic keys, or user data during execution, unless the TEE-hardware key is compromised---this could be possible if and only if an attacker physically de-capsulate chips, probe internal buses, or defeat the CPU's hardware security mechanisms.

\subsubsection{Data Integrity and Code Authenticity} 
\leavevmode\\ 
Remote attestation provides cryptographic evidence regarding the identity and integrity of the executing workload and its codebase. Confidential containers platforms verifies attestation evidence before provisioning cryptographic materials, ensuring that only approved software configurations receive access to sensitive credentials. In combination with the decentralized code management, this mechanism mitigates attacks involving unauthorized container images, modified runtimes, or malicious workload substitutions.
 Any deviation in the execution environment invalidates key release conditions enforced by KMS nodes and KMS smart contract.

\subsubsection{Collusion Resistance and Cryptographic keys Exfiltration/leakage} 
\leavevmode\\ 
An adversary must compromise at least $t$ out of $n$ key management nodes to reconstruct the seed. Compromise of fewer nodes yields no information about the secret due to the information-theoretic security of SSS. Further, cryptographic material is provisioned only after successful remote attestation of the confidential container. Secrets are released exclusively to SSI-Wallet workloads executing inside verified TEEs according to the authorized codebase, preventing unauthorized access by compromised hosts or orchestration components. Since private keys remain within the attested execution environment, usual key leakage attack vectors based on filesystem access, memory inspection, or privileged host access are  mitigated.

\subsubsection{Hardware Limitations} 
\leavevmode\\
The security of the architecture ultimately rely on the trust associated to  the TEE implementation and the correct enforcement of remote attestation policies. Compromise of the TEE hardware or its attestation mechanism would violate the confidentiality guarantees of the system. The unavailability of the vendor's certificate verification endpoint can cause a denial of service, as verifies can no longer check the validity of remote attestations.

\section{Conclusions}
\label{Conclusion}

Self-Sovereign Identity (SSI) is approaching a stage where it is increasingly recognized as a standard for digital identity management and is becoming technologically feasible for widespread adoption. Emerging technologies such as Distributed Ledger Technologies (DLTs), along with privacy-enhancing mechanisms like Trusted Execution Environments (TEEs) and zero-knowledge proofs (ZK proofs), provide strong opportunities to build SSI-compliant systems. However, technological choices influenced primarily by regulatory interpretations may hinder the full realization and recognition of SSI. This is evident in the current Italian and European Digital Identity Wallet, where centralized design decisions prioritize privacy through legal accountability rather than cryptographic transparency and technical verifiability.

This article presents an SSI-compliant wallet architecture that enables centralized verification of EUDI credentials without violating SSI principles. The proposed system relies on workloads executed within TEEs, with a publicly available and reproducible codebase. It incorporates decentralized code management through peer-to-peer confidential Key Management Service (KMS) nodes and a permissioned hybrid blockchain network. Remote attestations generated by TEEs are used to ensure correct workload execution, identify authorized TEE instances, and support a threshold secret sharing protocol that releases cryptographic material only to verified TEE instances.

In terms of future research directions, credential revocation via decentralized mechanisms remains superior to current centralized approaches. Hybrid solutions that are compatible with both decentralized applications and traditional systems show strong potential, such as Merkle tree-based revocation registries \cite{sitouah2024untraceable, munoz2004certificate}, on-chain commitments \cite{sitouah2024blockchain}, and event-log-based systems \cite{chatzigiannis2023sok}.

%Because tables cannot be split across pages, the best placement for them is typically the top of the page nearest their initial cite.  

\bibliographystyle{ACM-Reference-Format}
 
\bibliography{name}

\appendix

\end{document}